\documentclass[a4paper,12pt]{article}

\begin{document}
\makeatletter
\renewcommand{\theequation}{\thesection.\arabic{equation}}
\@addtoreset{equation}{section}
\makeatother

\title{Kinematical Holographic Relation in Yang's Quantized Space-Time and Area-Entropy Relation in  $D_0$ Brane Gas System}

\author{\large Sho Tanaka\footnote{Em. Professor of Kyoto University, E-mail: st-desc@kyoto.zaq.ne.jp }}

\date{}
\maketitle

\vspace{-10cm}
\rightline{}
\vspace{12cm}
\thispagestyle{empty}

\centerline {\bf Abstract}
\vspace{0.5cm}

We investigate the possible inner relation between the familiar Bekenstein-Hawking's area-entropy relation and ours presented in a simple $D_0$ brane gas system on the basis of the kinematical holographic  relation [KHR], which was found in the Yang's quantized space-time subject to the noncommutative geometry. We find out that the relation between them is well understood through the idea of {\it elementary} Schwarzschild black hole realized on a single $[site]$ of Planckian scale in our scheme. Related arguments explain the origin of a certain kind of universality as seen in $\eta =1/4$ in the Bekenstein-Hawking relation, showing that $\eta$ is closely related to the extremal entropy of  the {\it elementary} Schwarzschild black hole.

\vspace{0.5cm}
 
{\bf Key words} :Bekenstein-Hawking's area-entropy relation; Noncommutative geometry; Kinematical holographic relation in Yang's quantized space-time; Elementary Schwarzschild black hole.

\newpage

\section{\normalsize Introduction}

The present paper is a full version of the preceding two serial papers, [1] and [2]. In the latter two papers, 
we have presented a new area-entropy relation [AER] in a simple $D_0$ brane gas system on the basis of the kinematical 
holographic relation [KHR]$^{[3]}$ found in the Yang's quantized space-time algebra (YSTA),$^{[4]-[5],[1]-[3]}$  and 
tried to make clear its possible inner relation to the ordinary Bekenstein-Hawking area-entropy relation.$^{[6]-[12]}$

Indeed, the arguments presented there is essentially based on the framework of [KHR], a kind of holographic relation in 
the Lorentz-covariant Yang's quantized space-time algebra(YSTA), which we called the kinematical holographic 
relation [KHR].$^{[3]}$ It was simply expressed as
\begin{eqnarray} 
\hspace{-3cm} [KHR] \hspace{2cm}       n^L_{\rm dof} (V_d^L)= {\cal A} (V_d^L) / G_d,
\end{eqnarray}
that is, the proportional relation between $n_{\rm dof}(V_d^L)$ and ${\cal A} (V_d^L)$ with proportional constant $G_d$
(see section 3), where $n_{\rm dof} (V_d^L)$ and ${\cal A}(V_d^L)$, respectively, denote the number of degrees of freedom 
of any $d$ dimensional bounded spatial region with radius $L$, $V_d^L$, in Yang's quantized space-time and the boundary 
area of the latter region in unit of $\lambda$ or Planck length.

In this connection, the recent remarks given by Strominger in ref. [12], "Five Problems in Quantum Gravity," 
are noticeable. The author points out the first problem, "Universality of the Bekenstein-Hawking area-entropy 
law"
\begin{eqnarray}
S_{BH} = {Area \over {4 \hbar G}},   
\end{eqnarray}
by saying that the entropy was accounted for, consistently with the generalized laws of thermodynamics, including 
the $ 1/4 $ prefactor, for certain five-dimensional supersymmetric black holes in string theory by counting some kind 
of quantum microstates. At the same time, the author importantly remarks: "However, in this construction and 
its generalizations, 
the fact that the entropy is proportional to the area comes out only at the last step of a long computation. It is not 
obvious why this should always turn out to be so. A simple universal relation like (1.2) demands a simple universal 
explanation. The problem is to find it."

Indeed, we expect that our kinematical holographic relation $[KHR]$ presented in (1.1) may be able to respond to {\it 
a simple universal explanation} demanded by the above author.  As was emphasized in [3], the relation [KHR] (1.1) 
essentially reflects the fundamental nature of the noncommutative 
geometry of YSTA, that is, a definite {\it  kinematical reduction of spatial degrees of freedom} in comparison with 
the ordinary lattice space.  It should be noted that the kinematical reduction of spatial degrees of freedom is expected 
to hold almost in the recent attempts of noncommutative space-time, which was started by the the pioneer work of 
Snyder's quantized space-time$^{[13],[14] }$ and immediately after extended to the Yang's one mentioned above$^{[4],[5]}$: 
The possibility of the kinematical reduction of spatial degrees of freedom may be well understood intuitively, 
in terms of the familiar {\it quantum correlation} among different components of spatial coordinates subject to 
the noncommutative relations. It presents a possibility of giving a simple clue to resolve the long-pending problem 
encountered in the Bekenstein-Hawking area-entropy relation$^{[6]-[12]}$ or the holographic principle,$^{[15]-[19]}$ 
that is, the apparent gap between the degrees of freedom of any bounded spatial region associated with entropy and 
of its boundary area.

 On the basis of this $[KHR]$ (1.1) mentioned above, we derived in refs. [1] and [2] a new 
area-entropy relation of the $D_0$ brane gas system$^{[20]-[24]}$ constructed on YSTA according to the idea of 
M-theory$^{[25]}$: 
\begin{eqnarray}
\hspace{-3cm} [AER] \hspace{2cm}  S(V_d^L) \leq {\cal A} (V_d^L) {S_S[site] \over G_d}.
\end{eqnarray}
In the above expression, $S(V_d^L)$ on the left-hand side denotes the total entropy of the gas system inside $V_d^L$, 
which is expressed in our scheme in terms of the entropy of {\it individual} [site] with Planckian scale, 
$S[site]$, as $S(V_d^L) = n_{dof}(V_d^L) S[site].$\   $S_S [site]$ on the right-hand side describes the extremal value of  
$S[site]$, when the whole system tends to the Schwarzschild black hole, as will be reviewed in section 4. Then, one finds 
that the familiar Bekenstein proportional constant $\eta$ is now given by
 \begin{eqnarray}
\hspace{-3cm} \eta = {S_S[site] \over G_d}.
\end{eqnarray}

As was emphasized repeatedly in refs. [26], [1]-[3], YSTA which is intrinsically equipped with short- and long-scale 
parameters, $\lambda (= l_P)$ and $R$, gives a finite number of spatial degrees of freedom for any finite spatial region 
and provides a basis for the field theory free from ultraviolet- and infrared-divergences$^{[26]}$. In this line of thought, 
in what follows, we will put forward the arguments in refs. [1] and [2], as roughly explained above, and make clear 
the possible inner relation between the familiar Bekenstein-Hawking area-entropy relation (1.2) and our relation given 
in (1.3).

The present paper is organized as follows. In order to make the present paper as self-contained as possible, we first 
recapitulate Yang's quantized space-time algebra (YSTA) and its representations$^{[26],[1]-[3]}$ in section 2. 
Section 3 is devoted to review the derivation of the kinematical holographic relation [KHR] (in subsection 3.1) and to 
its extension to the lower-dimensional bounded regions, $V_d^L$ (in subsection 3.2).
 
In section 4, subsections 4.1 and 4.2 are devoted to review the main results in refs. [1]and [2]: In subsection 4.1, 
we first review a simple $D_0$ brane gas model and its statistical consideration through the entropy of the system 
constructed on the basic concept of [site] characteristic of noncommutative YSTA. In subsection 4.2, we derive a new 
area-entropy relation in the system in connection with Schwarzschild black hole, as was shown in (1.3). 
In subsection 4.3, we present the basic arguments on the physical implication of Hawking radiation temperature 
$T_{H.R.} (= 1/(8\pi M_S)$ and Bekenstein-Hawking relation with $\eta =1/4$ in a close connection with our scheme. 
Based on these arguments, in the last section 5, we attempt to identify $S_S[site]$ in (1.3) with the entropy of 
an {\it elementary} Schwarzschild black hole realized on a single $[site]$ of Planckian scale and further find out 
an important fact that the effective mass of $D_0$ brane inside black holes, $\mu_S$, is inversely proportional to 
the black hole mass $M_S$, almost in accord with the Hawking radiation temperature, $1/(8\pi M_S).$ Motivated by this 
fact, we attempt to examine further the physical implication of $\eta$ by introducing in our scheme an ansatz which 
enables us self-consistently to equate $\mu_S$ to $T_{H.R.}$ and leads us to $\eta$ slightly shifted from $1/4.$

\section{\normalsize Yang's Quantized Space-Time Algebra (YSTA) and \break Its Representations}  

\subsection{\normalsize Yang's Quantized Space-Time Algebra (YSTA) }

Let us first briefly review the Lorentz-covariant Yang's quantized space-time algebra (YSTA).
$D$-dimensional Yang's quantized space-time algebra is introduced$^{[1]-[3], [4]-[5]}$ as the 
result of the so-called Inonu-Wigner's contraction procedure with two contraction parameters, $R$ and 
$\lambda$, from $SO(D+1,1)$ algebra with generators $\hat{\Sigma}_{MN}$; 
\begin{eqnarray}
 \hat{\Sigma}_{MN}  \equiv i (q_M \partial /{\partial{q_N}}-q_N\partial/{\partial{q_M}}),
\end{eqnarray}
which work on $(D+2)$-dimensional parameter space  $q_M$ ($M= \mu,a,b)$ satisfying  
\begin{eqnarray}
             - q_0^2 + q_1^2 + \cdots + q_{D-1}^2 + q_a^2 + q_b^2 = R^2.
\end{eqnarray}
 
Here, $q_0 =-i q_D$ and $M = a, b$ denote two extra dimensions with space-like metric signature.

$D$-dimensional space-time and momentum operators, $\hat{X}_\mu$ and $\hat{P}_\mu$, 
with $\mu =1,2,\cdots,D,$ are defined in parallel by
\begin{eqnarray}
&&\hat{X}_\mu \equiv \lambda\ \hat{\Sigma}_{\mu a}
\\
&&\hat{P}_\mu \equiv \hbar /R \ \hat{\Sigma}_{\mu b},   
\end{eqnarray}
together with $D$-dimensional angular momentum operator $\hat{M}_{\mu \nu}$
\begin{eqnarray}
   \hat{M}_{\mu \nu} \equiv \hbar \hat{\Sigma}_{\mu \nu}
\end{eqnarray} 
and the so-called reciprocity operator
\begin{eqnarray}
    \hat{N}\equiv \lambda /R\ \hat{\Sigma}_{ab}.
\end{eqnarray}
Operators  $( \hat{X}_\mu, \hat{P}_\mu, \hat{M}_{\mu \nu}, \hat{N} )$ defined above 
satisfy the so-called contracted algebra of the original $SO(D+1,1)$, or Yang's 
space-time algebra (YSTA):
\begin{eqnarray}
&&[ \hat{X}_\mu, \hat{X}_\nu ] = - i \lambda^2/\hbar \hat{M}_{\mu \nu}
\\
&&[\hat{P}_\mu,\hat{P}_\nu ] = - i\hbar / R^2\ \hat{M}_{\mu \nu}
\\
&&[\hat{X}_\mu, \hat{P}_\nu ] = - i \hbar \hat{N} \delta_{\mu \nu}
\\
&&[ \hat{N}, \hat{X}_\mu ] = - i \lambda^2 /\hbar  \hat{P}_\mu
\\
&&[ \hat{N}, \hat{P}_\mu ] =  i \hbar/ R^2\ \hat{X}_\mu,
\end{eqnarray}
with other familiar relations concerning  ${\hat M}_{\mu \nu}$'s omitted.

\subsection{Quasi-Regular Representation of YSTA}

Let us further recapitulate briefly the representation$^{[26],[27]}$ of YSTA for the subsequent 
consideration in section 4. First, it is important to notice the following elementary fact that ${\hat\Sigma}_{MN}$ 
defined in Eq.(2.1) with $M, N$ being the same metric signature have discrete eigenvalues, i.e., $0,\pm 1 ,
\pm 2,\cdots$, and those with $M, N$ being opposite metric signature have continuous eigenvalues,
$\footnote{The corresponding eigenfunctions are explicitly given in ref. [27].}$ consistently with 
covariant commutation relations of YSTA. This fact was first emphasized by Yang$^{[4],[5]}$ with respect to 
the Snyder's quantized space-time mentioned above.$^{[13],[14]}$ This conspicuous aspect is well understood by means of 
the familiar example of the three-dimensional angular momentum in quantum mechanics, where individual components, 
which are noncommutative among themselves, are able to have discrete eigenvalues, consistently with the 
three-dimensional rotation-invariance. 
 
This fact implies that Yang's space-time algebra (YSTA) presupposes for its representation space 
to take representation bases like 
\begin{eqnarray}
| t/\lambda,n_{12}, \cdots> \equiv |{\hat{\Sigma}}_{0a} =t/\lambda> |{\hat{\Sigma}}_{12}=n_{12}>
\cdots|{\hat{\Sigma}}_{910}=n_{910}>,
\end{eqnarray}
where $t$ denotes {\it time}, the continuous eigenvalue of $\hat{X}_0 \equiv \lambda\ \hat{\Sigma}_{0 a}$ 
and $n_{12}, \cdots$ discrete eigenvalues of maximal commuting set of subalgebra of $SO(D+1,1)$ which are 
commutative with ${\hat{\Sigma}}_{0a}$, for instance, ${\hat{\Sigma}}_{12}$, ${\hat{\Sigma}}_{34},\cdots , 
{\hat{\Sigma}}_{910}$, when $D=11$.$^{[3],[26],[27]-[28]}$

Indeed, an infinite dimensional linear space expanded by $|\ t/\lambda, 
n_{12},\cdots>$ mentioned above provides a representation space of unitary infinite dimensional representation of YSTA. 
It is the so-called "quasi-regular representation"$^{[29]}$ of SO(D+1,1),\footnote{It corresponds, in the case of 
unitary representation of Lorentz group $SO(3,1)$, to taking $K_3\ (\sim \Sigma_{03})$ and $J_3\ (\sim \Sigma_{12})$ 
to be diagonal, which have continuous and discrete eigenvalues, respectively, instead of ${\bf J}^2$ and $J_3$ in 
the familiar representation.}
and is decomposed into the infinite series of the ordinary unitary irreducible representations of 
$SO(D+1,1)$ constructed on its maximal compact subalgebra, $SO(D+1)$. ( See  Chapter {\bf 10 },  10.1. "Decompositions of Quasi-Regular Representations and Integral Transforms" in ref. [29]. ) 

It means that there holds the following form of decomposition theorem,
\begin{eqnarray}
| t/\lambda, n_{12},\cdots>= \sum_{\sigma 's}\ \sum_{l,m}\  C^{\sigma's, n_{12}, \cdots }_{l,m}(t/\lambda)\ 
 | \sigma 's ; l,m>,
\end{eqnarray}      
with expansion coefficients $C^{\sigma's, n_{12}, \cdots}_{l,m}(t/\lambda).^{[3],[26],[27]}$ In Eq.(2.13), 
$|\sigma 's ; l, m>'s$ on the right hand side describe the familiar unitary irreducible representation 
bases of $SO(D+1,1)$, which are designated by $\sigma 's$ and $(l,m),$  
\footnote{In the familiar unitary irreducible representation of $SO(3,1)$, it is well known that $\sigma$'s are 
represented by two parameters, $(j_0, \kappa)$, with $j_0$ being $1,2, \cdots \infty$ and $\kappa$ being purely 
imaginary number, for the so-called principal series of representation. With respect to the associated representation 
of $SO(3)$, when it is realized on $S^2$, as in the present case, $l$'s denote positive integers, 
$l= j_0, j_0+1, j_0+2,\cdots,\infty$, and $m$ ranges over $\pm l, \pm(l-1) , \cdots,\pm1, 0.$ } 
denoting, respectively, the irreducible unitary representations of $SO(D+1,1)$ and the associated 
irreducible representation bases of $SO(D+1)$, the maximal compact subalgebra of $SO(D+1,1)$, mentioned above. 
It should be noted here that, as remarked in [26], $l$'s  are limited to be integer, excluding the possibility of 
half-integer, because of the fact that generators of $SO(D+1)$ in YSTA are defined as differential operators 
on $S^D$, i.e., ${q_1}^2 + {q_2}^2 + \cdots + {q_{D-1}}^2 + {q_a}^2 + {q_b}^2 = 1.$

In what follows, let us call the infinite dimensional representation space introduced above for the representation of 
YSTA, Hilbert space I, in distinction to Hilbert space II which is Fock-space constructed dynamically by 
creation-annihilation operators of second-quantized fields on YSTA, such as $D_0$ brane field discussed in section 4.

\section{\normalsize Kinematical Holographic Relation [KHR] in YSTA}

\subsection{\normalsize Recapitulation of Kinematical Holographic Relation [KHR]} 

First, let us remember that the following kinematical holographic relation\footnote{The argument in this 
subsection was given in ref. [3], based on refs.[27] and [28], in the following form of $D_0$ brane field equation: 
$[({X_\sigma}^2 +R^2 N^2 )( (\partial/\partial {X_\mu})^2 + R^{-2} (\partial/\partial {N})^2)) 
 - ( X_\mu \partial/\partial {X_\mu} + N \partial/\partial{N})^2 - (D-1)( X_\mu \partial/\partial {X_\mu}+N
 \partial/\partial{N})\ ]\ D ( X_\nu, N) = 0.$  Indeed, it was derived in refs.[28] from the following $D_0$ brane field 
action after M-theory,$^{[25]}$ 
$ \bar{\hat L} = A\ {\rm tr}\  \{ [\hat {\Sigma}_{KL}, 
\hat {D}^\dagger]\ [\hat {\Sigma}_{KL}, \hat {D}]\}=  A'\ {\rm tr}\  \{ 2\ (R^2 /\hbar^2)\  [{\hat P}_\mu, 
\hat {D}^\dagger ]\ [ \hat {P}_\mu, \hat {D}] - {\lambda}^{-4 }\ [\ [\hat {X}_\mu, \hat {X}_\nu], \hat {D}^\dagger] 
[ [\hat {X}_\mu,\hat {X}_\nu], \hat {D}]\},$ with $K, L = (\mu, b),$ by means of the Moyal star product method.} 
\begin{eqnarray}
\hspace{-3cm} [KHR] \hspace{2cm}       n^L_{\rm dof}= {\cal A} / G,
\end{eqnarray}
with the proportional constant $G$
\begin{eqnarray}
G\  \sim {(2 \pi)^{D/2} \over 2}\ (D-1)!! &&for\ D\ even
\\
    \sim  (2 \pi)^{(D-1)/2}(D-1)!! &&for\ D\ odd,
\end{eqnarray}
was derived in [3] for the $D$-dimensional space-like region with finite radius $L$ in D-dimensional 
Yang's quantized space-time in the unit of $\lambda$. Let us denote the region hereafter as $V_D^L$, 
which was defined by

\begin{eqnarray}
 \sum_{K \neq 0}{\Sigma_{aK}}^2  = \sum_{\mu \neq 0}{\Sigma_{a \mu}}^2 + {\Sigma_{ab}}^2 = (L/\lambda)^2,
\end{eqnarray}
or
\begin{eqnarray}
{X_1}^2 + {X_2}^2 + \cdots + {X_{D-1}}^2 + R^2\ N^2 = L^2.
\end{eqnarray}
Here, $\Sigma_{MN}$'s are presumed to be given in terms of Moyal star product formalism applied to the 
expression, $\Sigma_{MN}= ( -q_M p_n + q_N p_M)$, as was  treated in detail in [3].

${\cal A}$ in\ [KHR] (3.1) simply denotes the boundary surface area of $V_D^L$, that is,  
\begin{eqnarray}
{\cal A} =  ({\rm area\ of}\ S^{D-1}) ={(2 \pi)^{D/2} \over {(D-2)!!}} (L/\lambda)^{D-1} 
&&for\ D\ even,
\nonumber\\                   =2 {(2\pi)^{(D-1)/2} \over {(D-2)!!}} (L/\lambda)^{D-1} &&for\ D\ odd. 
\end{eqnarray}

On the other hand, $n^L_{\rm dof}$ in [KHR] (3.1), which denotes, by definition, the number of spatial 
degrees of freedom of YSTA inside $V_D^L$, was given in [3] as follows,  
\begin{eqnarray}
&&n^L_{\rm dof}  =  dim\ ( \rho_{[L/\lambda]}) = {2 \over (D-1)!}{([L/\lambda]+ D-2)! \over ([L/\lambda]-1)!}
\nonumber\\
 &&\hskip3.5cm \sim {2 \over (D-1)!}  [L/\lambda]^{D-1},
\end{eqnarray}
taking into consideration the fact $[L/\lambda] \gg D.$ 

Indeed, the derivation of the above equation (3.7) was the central task in [3]. In fact, we emphasized that 
the number of degrees of freedom $n^L_{\rm dof}$ inside $V_D^L$, which is subject to noncommutative algebra, YSTA, 
should be, logically and also practically, found in the structure of representation space of YSTA, that is, 
Hilbert space I defined in section 2. Let us here recapitulate in detail the essence of the derivation in order to 
make the present paper as self-contained as possible. 
   
In fact, one finds that the representation space needed to calculate $n^L_{\rm dof}$ is 
prepared in Eq.(2.13), where any "quasi-regular" representation basis,\\ $ | t/\lambda, n_{12}, \cdots>$, 
is decomposed into the infinite series of the ordinary unitary representation bases of 
$SO(D+1,1)$, $| \sigma 's ; l,m>.$ 
As was stated in subsection 2.2, the latter representation bases, $| \sigma 's ; l,m>'s$ are constructed on 
the familiar finite dimensional representations of maximal compact subalgebra of YSTA, $SO(D+1)$, whose representation 
bases are labeled by $(l,m)$ and provide the representation bases for spatial quantities under consideration, 
because $SO(D+1)$ just involves those spatial operators $( \hat{X}_u, R \hat{N})$. 

In order to arrive at the final goal of counting $n^L_{\rm dof}$, therefore, one has only to find mathematically 
a certain irreducible representation of $SO(D+1)$, which {\it properly} describes (as seen in what follows) 
the spatial quantities $( \hat{X}_u, R \hat{N})$ inside the bounded region with radius $L$, then one finds 
$n^L_{\rm dof}$ through counting the dimension of the representation. 

At this point, it is important to note that, as was remarked in advance in subsection 2.2, any generators of $SO(D+1)$ 
in YSTA  are defined by the differential operators on the $D-$dimensional unit sphere, $S^D$, i.e., 
${q_1}^2 + {q_2}^2 + \cdots + {q_{D-1}}^2 + {q_a}^2 + {q_b}^2 = 1,$ limiting its representations with $l$ to be 
integer.
   
On the other hand, it is well known that the irreducible representation of arbitrary high-dimensional $SO(D+1)$ 
on $S^D = SO(D+1)/SO(D)$ is derived in the algebraic way, $^{[30]}$( see Chapter II. \$5 "Riemann Symmetric Pairs"\\ and\  Chapter III. \$12 "Spherical Functions on Spheres" in ref. [30] ),  irrelevantly to any detailed knowledge of 
the decomposition equation (2.13), but solely in accord with the fact that $SO(D+1)$ in YSTA is defined originally 
on $S^D$, as mentioned above. One can choose, for instance, $SO(D)$ with generators $\hat{\Sigma}_{MN} (M,N=b, u)$, 
while $SO(D+1)$ with generators $\hat{\Sigma}_{MN}(M,N=a,b,u),$ where $u = 1,2,\cdots, D-1$. Then, it turns out that any irreducible representation 
of $SO(D+1)$, denoted by $\rho_l$, is uniquely designated by the maximal integer $l$ of eigenvalues of 
${\hat \Sigma}_{ab}$ in the representation, where ${\hat \Sigma}_{ab}$ is known to be a possible Cartan subalgebra 
of the so-called compact symmetric pair $(SO(D+1),SO(D))$ of rank $1$.

According to the so-called Weyl's dimension formula, the dimension of $\rho_l$ is given by$^{[30],[3],[26]}$
\begin{eqnarray}
 dim\ (\rho_l)= {(l+\nu) \over \nu} {(l+2\nu-1)! \over {l!(2\nu -1)!}},
\end{eqnarray}
where $ \nu \equiv (D-1)/2$ and $D \geq 2$. \footnote{This equation just gives the familiar result $dim\ (\rho_l)= 2l +1,$ 
in the case $SO(3)$ taking $D=2.$} ( See Eq. (12.5) Chapter III. \$12 "Spherical Functions on Spheres" in ref. [30].)

Finally, we can find a certain irreducible representation of $SO(D+1)$ among those $\rho_l 's $  given above, 
which {\it properly} describes (or realizes) the spatial quantities inside the bounded region $V_D^L$. 
Now, let us choose tentatively $l = [L/\lambda]$ with $[L/\lambda]$ being the integer part of $L/\lambda$. In this case, 
one finds out that the representation $\rho_{[L/\lambda]}$ just {\it properly} describes all of generators of 
$SO(D+1)$ inside the above bounded spatial region $V_D^L$, because  $[L/\lambda ]$ indicates also the largest eigenvalue 
of any generators of $SO(D+1)$ in the representation $\rho_{[L/\lambda]}$ on account of its $SO(D+1)-$invariance and 
hence eigenvalues of spatial quantities $( \hat{X}_u, R \hat{N})$ are well confined inside the bounded region with 
radius {L}. As the result, one finds that the dimension of $\rho_{[L/\lambda]}$ just gives the number of spatial degrees 
of freedom inside $V_D^L$, $n^L_{\rm dof}$, as shown in (3.7) and thus the relation [KHR] (3.1) 
holds with $G$ given by (3.2) and (3.3).

\subsection{\normalsize KHR in the lower-dimensional spatial region $V_d^L$ }
According to the argument given for $V_D^L$ in the preceding subsection, let us study the 
kinematical holographic relation in the lower-dimensional bounded spatial region $V_d^L$ for the subsequent 
argument of the area-entropy relation in section 4. In fact, it will be given through a simple $D_0$ brane gas system 
formed inside $d\ (\leq{D-1})$-dimensional bounded spatial region, $V_d^L$, which is defined 
by
\begin{eqnarray}  
{X_1}^2 + {X_2}^2 + \cdots + {X_d}^2 = L^2,
\end{eqnarray}
instead of (3.5).

In this case, the boundary area of $V_d^L$, that is, ${\cal A}\ (V_d^L)$ is given by
\begin{eqnarray}
{\cal A}\ (V_d^L) =  ({\rm area\ of}\ S^{d-1}) ={(2 \pi)^{d/2} \over {(d-2)!!}} (L/\lambda)^{d-1} 
&&for\ d\ even
\nonumber\\                   =2 {(2\pi)^{(d-1)/2} \over {(d-2)!!}} (L/\lambda)^{d-1} &&for\ d\ odd, 
\end{eqnarray}
corresponding to Eq. (3.6).

On the other hand, the number of degrees of freedom of $V_d^L$, let us denote it $n_{dof} (V_d^L)$, is calculated 
by applying the arguments given for derivation of $n_{dof}^L$ in (3.7). In fact, it is found in a certain irreducible 
representation of $SO(d+1)$, a minimum subalgebra of YSTA, which includes the $d$ spatial quantities, $\hat{X}_1, 
\hat{X}_2, \cdots, \hat{X}_d$ needed to properly describe $V_d^L$, and is really constructed by the generators 
$\hat{\Sigma}_{MN}$ with $M,N$ ranging over $a,1,2, \cdots,d$. The representation of $SO(d+1)$, let us denote it 
$\rho_l\ (V_d^L)$ with suitable integer $l = [L/\lambda]$, is given on the representation space $S^d =SO(d+1)/SO(d)$, 
taking the subalgebra $SO(d)$, for instance, $\hat{\Sigma}_{MN}$ with $M,N$ ranging over 
$1,2,\cdots,d$, entirely in accord with the argument on the irreducible representation of $SO(D+1)$ given in the 
preceding subsection 3.1. 

One immediately finds that 
\begin{eqnarray}
&&n_{\rm dof}\ (V_d^L)  =  dim\ ( \rho_{[L/\lambda]}\ (V_d^L)) = {2 \over (d-1)!}{([L/\lambda]+ d-2)! \over ([L/\lambda]-1)!}
\nonumber\\
 &&\hskip3.5cm \sim {2 \over (d-1)!}  [L/\lambda]^{d-1}
\end{eqnarray}
corresponding to (3.7), and there holds, from (3.10) and (3.11), the following kinematical holographic 
relation for $V_d^L$ in general
\begin{eqnarray}
\hspace{-3cm} [KHR] \hspace{2cm}       n_{\rm dof}\ (V_d^L)=
{\cal A}\ (V_d^L) / G_d,
\end{eqnarray}
with the proportional constant $G_d$
\begin{eqnarray}
G_d \sim {(2 \pi)^{d/2} \over 2}\ (d-1)!! &&for\ d\ even
\\
    \sim  (2 \pi)^{(d-1)/2}(d-1)!! &&for\ d\ odd,
\end{eqnarray}
corresponding to Eqs. (3.1)- (3.3) for $V_D^L$.  
     
\section{\normalsize Area-Entropy Relation in $D_0$ Brane Gas System subject to YSTA}
  
\subsection{\normalsize $D_0$ Brane Gas Model in $V_d^L$ and Its Mass and Entropy}
Now, let us consider the central problem of the present paper, that is, the derivation of a possible area-entropy relation 
through a simple $D_0$ brane gas$^{[20]-[24]}$ model formed inside $V_d^L$ according to the idea of M-theory.$^{[25]}$ 
This possibly implies that one has to deal with the dynamical system of the second-quantized $D_0$ brane field 
${\hat D}_0$ inside $V_d^L$. In the present toy model of the $D_0$ brane gas, however, we avoid to enter into detail 
of the dynamics of $D_0$ brane system, but treat it as an ideal gas, only taking into consideration 
that the system is developed on $V_d^L$ subject to YSTA and its representation discussed above, but neglecting 
interactions of $D_0$ branes, for instance, with strings as well as self-interactions among themselves. 

First of all, according to the argument given in the preceding subsection 3.2, the spatial structure of $V_d^L$ 
is described through the specific representation $ \rho_{[L/\lambda]}\ (V_d^L)$. Let us denote its orthogonal 
basis-vector system in Hilbert space I, as follows 
\begin{eqnarray}
 \rho_{[L/\lambda]}\ (V_d^L): \quad |\ m >,  \qquad  m= 1,2,\cdots, n_{\rm dof}(V_d^L).     
\end{eqnarray}   
 In the above expression, $n_{\rm dof}\ (V_d^L)$ denotes the dimension of the representation 
$\rho_{[L/\lambda]}\ (V_d^L)$, as defined in (3.11).

 One notices that the labeling number $m$ of basis vectors, which ranges from $1$ to $n_{\rm dof}(V_d^L)$ 
plays the role of {\it spatial coordinates} of $V_d^L$ in the present noncommutative YSTA, corresponding to the so-called 
lattice point in the lattice theory. Let us denote the {\it point} hereafter $[site]$ or $[site\ m]$ of $V_d^L$.

At this point, one should notice that the {\it second quantized} ${\hat D}_0$-brane field$^{[31]-[32]}$ on $V_d^L$ 
must be the linear operators operating on Hilbert space I, and described by $n_{\rm dof}(V_d^L) \times n_{\rm dof}(V_d^L)$ 
matrix under the representation $\rho_{[L/\lambda]}\ (V_d^L)$ like $< m\ |{\hat D}|\ n >$ on the one hand, and 
on the other hand each matrix element must be operators operating on Hilbert space II, playing the role of 
creation-annihilation of $D_0$ branes. On the analogy of the ordinary quantized local field, let us define 
those creation-annihilation operators through the diagonal parts in the following way:\footnote{On the other hand, 
the non-diagonal parts, $< m\ |{\hat D}|\ n >,$ 
are to be described in terms like ${\bf a}_m {\bf a}_n^\dagger$ or ${\bf a}_m^\dagger {\bf a}_n$ in accord with 
the idea of M-theory where they are conjectured to be concerned with the interactions between $[site\ m]$ and $[site\ n].$ 
The details must be left to the rigorous study of the second quantization of $D_0$-brane field.$^{[32]}$}    
\begin{eqnarray}
< m\ | {\hat D}|\ m >\  \sim\ {\bf a}_m\ {\rm or}\  {\bf a}_m^\dagger.
\end{eqnarray}
In the above expression, ${\bf a}_m$ and ${\bf a}_m^\dagger$, respectively, denote annihilation and creation 
operators of $D_0$ brane, satisfying the familiar commutation relations,
\begin{eqnarray}
&&[{\bf a}_m, {\bf a}_n^\dagger]= \delta_{mn},
\\
&&[{\bf a}_m, {\bf a}_n]= 0 .
\end{eqnarray}

Now, let us focus our attention on quantum states constructed dynamically in Hilbert space II by the creation-annihilation 
operators ${\bf a}_m$ and ${\bf a}_m^\dagger$ of $D_0$ branes introduced above at each [site] inside $V_d^L$. One should 
notice here the important fact that in the present simple $D_0$ brane gas model neglecting all interactions of $D_0$ branes, 
each $[site]$ can be regarded as independent quantum system and described in general by own statistical operator, while 
the total system of gas is described by their direct product. In fact, the statistical operator at each $[site\  m]$ 
denoted by ${\hat W}[m]$, is given in the following form,
\begin{eqnarray} 
{\hat W}[m] = \sum_ k  w_k\  |\ [m]: k >\  < k :[m]\ |,
\end{eqnarray}
with 
\begin{eqnarray}
 |\ [m]:  k  >  \equiv {1 \over \sqrt{k!}}({\bf a}_m^\dagger)^{k}|\ [m]:0 >.
\end{eqnarray}
That is, $|\ [m]:  k > ( k=0, 1, \cdots)$ describes the normalized quantum-mechanical state in Hilbert space II with 
$k$ $D_0$ branes constructed by ${\bf a}_m^\dagger$ on $|\ [m]:0 >,$ i.e. the vacuum state of $[site\ m]$.
\footnote{The proper vacuum state in Hilbert space II is to be expressed by their direct product.} And $w_k$'s denote 
the realization probability of state with occupation number $k$, satisfying $\sum_k w_k = 1.$ 

We assume here that the statistical operator at each $[site\ m]$ is common to every [site] in the present $D_0$ 
brane gas under equilibrium state, with the common values of $w_k$'s and the statistical operator of total 
system on $V_d^L$ , ${\hat W}(V_d^L)$, is given by  
\begin{eqnarray}
{\hat W}(V_d^L) = {\hat W}[1] \otimes {\hat W}[2] \cdots \otimes {\hat W}[m] \cdots \otimes {\hat W}[n_{dof}].
\end{eqnarray}

Consequently, one finds that the entropy of the total system, $S(V_d^L)$, is given by
\begin{eqnarray}
S(V_d^L) = - {\rm Tr}\ [{\hat W}(V_d^L)\ {\rm ln} {\hat W}(V_d^L)] = n_{dof}(V_d^L)\times S[site], 
\end{eqnarray}
where $S[site]$ denotes the entropy of each [site] assumed here to be common to every [site] and given by
\begin{eqnarray}
S[site] = - {\rm Tr}\ [ {\hat W}[site]\ {\rm ln}{\hat W}[site]] = - \sum_k w_k\  {\rm ln} w_k.
\end{eqnarray}

Comparing this result (4.8) with [KHR]\ (3.12) derived in the preceding section, we find an important fact that the 
entropy $S(V_d^L)$ is proportional to the surface area ${\cal A}\ (V_d^L)$, that is, a kind of area-entropy relation 
([AER]) of the present system:
\begin{eqnarray}
\hspace{-3cm} [AER] \hspace{2cm}         S(V_d^L) = {\cal A}\ (V_d^L)\ {S[site] \over G_d},
\end{eqnarray}
where $G_d$ is given by (3.13)-(3.14). 

Next, let us introduce the total energy or mass of the system, $M(V_d^L)$. If one denotes the average energy or mass 
of the individual $D_0$ brane inside $V_d^L$ by $\mu$, it may be given by
\begin{eqnarray}   
M(V_d^L) = \mu {\bar N}[site]\  n_{\rm dof}(V_d^L) \sim  \mu {\bar N}[site] {2  \over (d-1)!}  [L/\lambda]^{d-1},
\end{eqnarray}
where ${\bar N}[site]$ denotes the average occupation number of $D_0$ brane at each $[site]$ given by
\begin{eqnarray}
{\bar N}[site] \equiv \sum_k k w_k.
\end{eqnarray}

Comparing this expression (4.11) with (4.8) and (3.12), respectively, we obtain a kind of mass-entropy relation ([MER]) 
\begin{eqnarray}
\hspace{-2cm} [MER] \hspace{2cm}         M(V_d^L) / S(V_d^L) = \mu {\bar N}[site] / S[site],
\end{eqnarray}
and  a kind of area-mass relation ([AMR])
\begin{eqnarray}
\hspace{-3cm} [AMR] \hspace{2cm} M(V_d^L) = {\cal A}(V_d^L)\ {\mu {\bar N}[site] \over G_d}.
\end{eqnarray}
     
\subsection{\normalsize Schwarzschild Black Hole and Area-Entropy Relation In $D_0$ brane Gas System}

In the preceding subsection 4.1, we have studied $D_0$ brane gas system and derived area-entropy relation $[AER]$ 
(4.10), mass-entropy and area-mass relations, $[MER]$ (4.13) and $[AMR]$ (4.14), which are essentially based on 
the kinematical holographic relation in YSTA studied in section 3. 

At this point, it is quite important to notice that these three relations explicitly depend on 
the following ${\it static}$ quantities of the gas system, $\mu$, ${\bar N}[site]$ and $S[site]$, that is, the average 
energy of individual $D_0$ brane, the average occupation number of $D_0$ branes and the entropy at each [site], 
which are assumed to be common to every [site], while these quantities turn out to play an important role in 
arriving finally at the area-entropy relation in connection with black holes, as will be seen below.   

Now, let us investigate how the present gas system tends to a black hole. We assume for simplicity that the system 
is under $d=3$, and becomes a Schwarzschild black hole, in which the above quantities acquire certain limiting 
values, $ \mu_S$, ${\bar N}_S[site]$ and $S_S [site]$, while the size of the system, $L$, becomes $R_S$, 
that is, the so-called Schwarzschild radius given by 
\begin{eqnarray}   
R_S =  2 G M_S(V_3^{R_S})/c^2,
\end{eqnarray}
where $G$ and $c$ denote Newton's constant and the light velocity, respectively, and $M_S(V_3^{R_S})$ is given by 
Eq. (4.11) with $L= R_S$, $\mu = \mu_S$ and ${\bar N}[site] = {\bar N}_S[site]$. Indeed, inserting 
the above values into Eq.(4.11), we arrive at the important relation, called hereafter the black hole condition [BHC], 
\begin{eqnarray}
\hspace{-1cm} [BHC] \hspace{2cm} M_S(V_3^{R_S}) = {\lambda^2 \over 4\mu_S {\bar N}_S [site]}{c^4 \over G^2} 
= {M_P^2 \over 4 \mu_S {\bar N}_S[site]}.
\end{eqnarray}
In the last expression, we assumed that $\lambda$, i.e., the small scale parameter in YSTA is equal  
to Planck length $l_P = [G \hbar / c^3]^{1/2} = \hbar /( c M_P )$, where $M_P$  denotes Planck mass. In what follows, 
we will use Planck units in $D=4$ or $d=3$, with $M_P = l_P = \hbar = c = k =1$.

Now, we simply obtain the area-entropy relation [AER] under the Schwarzschild black hole by 
inserting the above limiting values into [AER] (4.10)   
\begin{eqnarray}
S_S(V_3^{R_S}) = {\cal A}\ (V_3^{R_S})\ {S_S[site] \over 4 \pi},
\end{eqnarray}
noticing that $G_{d = 3} =4 \pi.$

At this point, one finds that it is a very important problem how to relate the above area-entropy relation 
under a Schwarzschild black hole with [AER] (4.10) of $D_0$ brane gas system in general, which is 
derived irrelevantly of the detail whether the system is a black hole or not. As was mentioned in the beginning 
of this subsection, however, the problem seems to exceed the applicability limit of the present toy model 
of $D_0$ brane gas, where the system is treated solely as a {\it static} state under {\it given} values of 
parameters, $\mu$, ${\bar N}[site]$ and $S[site]$, while the critical behavior around the formation of Schwarzschild 
black hole must be hidden in a possible ${\it dynamical}$ change of their values. 

In order to supplement such a defect of the present static toy model, let us try here a Gedanken-experiment, 
in which one increases the entropy of the gas system $S(V_3^L)$, keeping its size $L$ at the initial value $L_0$, 
until the system tends to a Schwarzschild black hole, where Eqs. (4.16) and (4.17) with $R_S = L_0$ hold. Then, one 
finds that according to [AER] (4.10), the entropy of $[site]$, $S[site]$ increases proportionally to $S(V_3^L)$ and 
reaches the limiting value $S_S [site]$, starting from any initial value $S_0[site]$ prior to formation of the black 
hole, because ${\cal A}(V_d^L)$ in Eq.(4.10) is invariant during the process. Namely, one finds a very simple fact 
that $S_0[site] \leq S_S [site].$  However, this simple fact combined with [AER] (4.10) leads us to the following 
form of a new area-entropy relation which holds throughout for the $D_0$ brane gas system up to the formation of 
Schwarzschild black hole,\footnote{Similarly, by the second Gedanken-experiment, in which one increases 
the total mass of gas system $M(V_3^L)$ with the fixed size $L_0$ in connection with [AMR] (4.14), 
in place of the increase of the entropy of gas system $S(V_3^L)$ in the first Gedanken-experiment, 
one obtains a new area-mass relation [AMR], $M(V_3^L) \leq {\cal A}(V_3^L) \mu_S {\bar N}_S[site] / 4 \pi$.}
\begin{eqnarray}
\hspace{-3cm} [AER] \hspace{2cm}  S(V_3^L) \leq {{\cal A} (V_3^L) S_S[site] \over 4 \pi},
\end{eqnarray}
where the equality holds for Schwarzschild black hole, as seen in Eq. (4.17).

\subsection{\normalsize The Possible Inner Relation between Our Approach and Bekenstein-Hawking Relation: Preliminaries}

We have derived the area-entropy relation [AER] (4.18) together with (4.10) in our toy model of $D_0$ brane gas system 
subject to Yang's quantized 
space-time algebra, YSTA. The relation [AER] (4.18) is to be compared with the original Bekenstein proposal
\begin{eqnarray}
S \leq \eta {\cal A},
\end{eqnarray}
where the proportional constant $\eta$ is now given in terms of a physical quantity, i.e., the partial entropy of 
the individual [site] of ${D_0}$ brane gas system under Schwarzschild black hole, that is, $S_S[site]$, as follows, 
\begin{eqnarray}
\eta = {S_S[site] \over 4 \pi}.
\end{eqnarray}

In addition, it is well-known that the Bekenstein proposal (4.19) was extended to the Bekenstein-Hawking Area-entropy 
relation 
\begin{eqnarray}
S \leq {\cal A} /4
\end{eqnarray}
with $\eta$ fixed to be $1/4$ through the investigation of the so-called Hawking radiation of black hole, 
which suggests to us more specifically that
\begin{eqnarray}
 S_S[site] = \pi.
\end{eqnarray}

In order to make clear the implication of the above {\it constraint} (4.20) or (4.22), let us supplement our preceding 
arguments by introducing the concept of temperature $T$ of the present gas system of $D_0$ branes, through the entropy 
of individual [site] mentioned above. It was assumed to be common to every $[site]$ of $D_0$ brane gas system in 
equilibrium, given by $S[site] =  {\rm Tr}\ [ {\hat W}[site]\ {\rm ln}{\hat W}[site]]= - \sum_k w_k\  {\rm ln} w_k $ 
in Eq. (4.9). 

Now let us take the following familiar expression for $w_k$'s,
\begin{eqnarray}
 w_k = e^{-\mu k/T} / Z(T)
\end{eqnarray}
where
\begin{eqnarray}
Z(T) \equiv \sum_{k=0}^\infty e^{-\mu k/T} = 1 / (1- e^{-\mu / T}). 
\end{eqnarray}

Then, one finds that 
\begin{eqnarray}
S[site]\equiv - \sum_k w_k\  {\rm \ln} w_k = - \ln (1 - e^{- \mu /T}) + {\mu \over T}\ ( e^{\mu /T}- 1)^{-1},
\end{eqnarray}
\begin{eqnarray}
{\bar N}[site] \equiv \sum_k k w_k = ( e^{\mu / T}- 1)^{-1},
\end{eqnarray}
and there holds the relation
\begin{eqnarray}
S [site] = \ln (1+ {\bar N}[site]) + {\bar N} [site] \ln (1+ {\bar N}^{-1}[site]),
\end{eqnarray}
where ${\bar N}[site]$ is the average occupation number of $D_0$ brane at each site, as defined in Eq. (4.12). 

At this point, let us apply the above result to our gas system under Schwarzschild black hole condition 
[BHC] considered in the preceding subsection 4.2, where all physical quantities were denoted with subscript $S$, 
such like $T_S$. First of all, one notices that Eq.(4.26) combined with Eq. (4.16) [BHC] 
\begin{eqnarray}
{\bar N}_S[site] (= ( e^{\mu_S /T_S}-1)^{-1}) = 1 / (4 \mu_S M_S)
\end{eqnarray}
gives rise to the important expressions for $T_S$ for the later discussion,   
\begin{eqnarray}
 T_S =& {\mu}_S / \ln (1 + {\bar N}^{-1}[site])
\nonumber  
\end{eqnarray}
\begin{eqnarray}
  =& {1 \over {4 M_S}} {\bar N}_S^{-1} / \ln(1+ {\bar N}_S^{-1}[site]).
\end{eqnarray} 

Furthermore, one finds that the relation (4.27) combined with Eq. (4.28) gives    
\begin{eqnarray}
S_S [site]\ ( = \ln (1+ {\bar N}_S[site]) + {\bar N}_S [site] \ln (1+ {\bar N}_S^{-1}[site]))
\nonumber\\ = \ln (1+ (4\mu_SM_S)^{-1}) + {1 \over {4 \mu_S M_S}} \ln (1 + 4\mu_S M_S). 
\end{eqnarray}

On the bases of the above relations, let us try in our present scheme to find the so-called 
Hawking radiation temperature $T_{H.R.}$ of the gas system under Schwarzschild black hole, which is defined by 
\begin{eqnarray}
T_{H.R.}^{-1} =  dS_S/dM_S.
\end{eqnarray}
Noticing the relation [AER] (4.17) with $ {\cal A} = 4 \pi R_S^2 =16 \pi M_S^2$, we immediately 
find 
\begin{eqnarray} 
T_{H.R.}^{-1} ={d \over dM_S} S_S  = {d \over d M_S} (16 \pi M_S^2 S_S[site]/{4 \pi})     \cr
\nonumber\\             = 8 M_S S_S[site] + 4M_S^2  {d \over d M_S}{S_S[site]}.
\end{eqnarray}

Now, in the calculation of the second term in the last expression of Eq. (4.32), it becomes important 
how to consider the possible dependence of $\mu_S$ on the total mass $M_S$ 
in $S_S[site]$, whose explicit expression is given in (4.30). 

Leaving its comprehensive discussion to the last section, let us take here an assumption that 
$S_S [site]$ (and hence ${\bar N}[site]$ on account of the relation (4.27)) is 
a certain universal constant independent of individual black holes with mass $M_S$, 
as in the cases of Bekenstein-Hawking constraints (4.20) ($ S_S[site] = 4 \pi \eta $) or (4.22) 
( $S_S[site] =  \pi $) with $\eta = 1/4$. Then, from Eq. (4.32) we simply obtain
\begin{eqnarray}
T_{H.R.}^{-1} = 8 M_S S_S[site] \hspace{8cm}
\nonumber \\   = 8 M_S \ \ln (1+ (4\mu_SM_S)^{-1}) + {2 \over { \mu_S}} \ln (1 + 4\mu_S M_S).   
\end{eqnarray}

Here, it is interesting especially to notice the latter case (4.22) corresponding to $\eta = 1/4$, 
that is,
\begin{eqnarray}
S_S[site] = \ln (1+ (4\mu_SM_S)^{-1}) + {1 \over {4 \mu_S M_S}} \ln (1 + 4\mu_S M_S)) = \pi.
\end{eqnarray} 
 It immediately leads us, through (4.33), to the well-known Hawking radiation temperature,  
\begin{eqnarray}
T_{H.R.} (= 1/(8 M_S S_S[site])) = {1 \over {8 \pi M_S}},    
\end{eqnarray}
and furthermore to the following important results that $\mu_S M_S $ and hence $\bar N_S[site] (= 1/(4\mu_S M_S)$, 
see (4.27)), i.e., the average occupation number of $D_0$ branes at each [site], respectively, takes  
a certain fixed value, that is,
\begin{eqnarray} 
\mu_S M_S (/M_P^2) \sim 0.03 
\end{eqnarray}
and 
\begin{eqnarray}
 \bar N_S[site] \sim 1/ 0.12, 
\end{eqnarray}
irrespectively of individual black holes.

Indeed, from Eqs. (4.35) and (4.36), one notices the {\it similarity} between the order of magnitudes of $T_{H.R.} 
(=1/(8\pi M_S)) \sim 0.04 /M_S$ and $\mu_S\sim 0.03/M_S$, that is, 
\begin{eqnarray}
T_{H.R.} \sim (0.04 / 0.03)\  \mu_S.
\end{eqnarray} 
In contrast, one finds in this case ( 4.37) 
\begin{eqnarray}
T_S \sim& (1 /0.11)\ \mu_S,
\end{eqnarray}
from (4.29).

With respect to the marked difference between $T_{H.R.}$ and $T_S$ as seen in Eqs. (4.38) and (4.39), 
one should notice that $T_S$ is introduced in a simple limit of the ordinary temperature $T$ of 
the statistical system of $D_0$ branes, at the moment when the system tends to Schwarzschild black hole, 
while the Hawking radiation temperature $T_{H.R.}$, as seen in (4.31), is defined through the $M_S$-variational 
process of Schwarzschild black hole of the $D_0$ branes system which is throughout subject to [BHC] (4.16). As seen in 
the derivation of $[BHC]$, which is given in connection with Schwarzschild radius $R_S$ given by Eq. (4.15), 
one finds out a certain kind of {\it nonlinear} character of $[BHC]$, which is evident in 
the relation (4.16), where $M_S (= M(V_3^{R_S}))$ is {\it inversely} proportional to $\mu_S {\bar N}_S[site]$ 
under $[BHC]$, while $M (V_3^L)$ in (4.11) is proportional to $\mu {\bar N}[site]$ in accord with the original 
definition of the effective mass of the ordinary statistical system of $D_0$ branes.  Thus, one should remark that Hawking 
radiation temperature $T_{H.R.}$ plays the role of a genuine temperature of the black hole system 
under $[BHC]$, giving the strong correlation with $\mu_S$, the effective mass of $D_0$ branes inside the black hole, 
as seen in (4.38), although $T$ or $T_S$ also plays the important role, for instance, in giving the relation between 
$ S[site]$ and $ {\bar N}[site]$ throughout before and after the formation of black hole, as seen in Eqs. (4.25) - (4.30).

We will further intensify the above arguments in the last section. 

\section{\normalsize Discussion and Concluding Remarks}

 In the present paper, we have started with the argument of the kinematical holographic relation [KHR] (1.1) 
found in the Lorentz-covariant Yang's quantized space-time algebra (YSTA). As was emphasized 
there, it essentially reflects the fundamental nature of the noncommutative geometry of YSTA itself, that is, a definite 
kinematical reduction of spatial degrees of freedom in comparison with the ordinary lattice space. Furthermore, 
YSTA, which is intrinsically equipped with short- and long-scale parameters, $\lambda (=l_P)$ and $R$, gives a finite 
number of spatial degrees of freedom for any finite spatial region and provides a basis for the field theory 
free from ultraviolet- and infrared-divergences.$^{[3],[26],[33]}$ 

Therefore, the argument on the holographic 
relation and area-entropy relation of $D_0$ brane gas system extended in the present paper has attached importance to 
{\it the first principle} of YSTA as much as possible, although it may be too crude and simple 
to treat physics around Planckian scale. 

From this point of view, the expression of our area-entropy relation [AER] (4.18) or (4.19) 
with $\eta = S_S[site]/4 \pi$ (4.20) must be one of our central concern. Indeed, the former relation (4.18) has 
been naturally derived from the kinematical holographic relation mentioned above and extensively studied in 
subsection 4.3 in connection with Bekenstein-Hawking relation and Hawking radiation temperature $T_{H.R.}$. We expect 
that it possibly gives us a clue 
to search for physics around the Planckian scale, from the standpoint that the Bekenstein-Hawking relation as well as 
Hawking radiation temperature provides us, not necessarily  {\it a priori} principle of black holes, but rather 
important empirical knowledge. From this point of view, we further scrutinize several results derived in subsection 4.3.

1. First, it is important to focus our attention on $S_S[site]$, on which we have pointed out in subsection 4.3 
a certain kind of {\it universality}, such like $S_S[site] = \pi\ (4.22)$ in connection with Bekenstein-Hawking 
relation or $\eta =1/4.$

Let us start with Eq. (4.17), which was given prior to [AER] (4.18):
\begin{eqnarray}
S_S(V_3^{R_S}) = {\cal A}\ (V_3^{R_S})\ {S_S[site] \over 4 \pi}.  
\end{eqnarray}
$S_S(V_3^{R_S})$ on the right-hand side is given by (4.8)
\begin{eqnarray}
S_S(V_3^{R_S}) = n_{dof}(V_3^{R_S}) S_S[site]   
\end{eqnarray}
with $R_S = 2M(V_3^{R_S})$ (4.15).

Now, let us consider an extreme case, $ R_S = \lambda\ ( = l_P)$ in (5.1) and (5.2). Then, taking into consideration that 
$ {\cal A} (V_3^{\lambda}) = 4 \pi$  and  $n_{dof}(V_3^\lambda) = 1$  according to (3.10) and (3.11), respectively, 
one finds that both Eqs. (5.1) and (5.2) lead us to the result,
\begin{eqnarray}
S_S(V_3^{R_S=l_P}) = S_S[site].
\end{eqnarray}

This result  tells us that $S_S[site]$ under consideration is nothing but the {\it extremal} entropy of an 
{\it elementary} black hole, $S_S(V_3^{R_S=l_P})$, which consists of a single [site] with Planckian scale and must 
have the mass $M_S(V_3^{R_S=l_P}) (= R_S /2$, see (4.15)) $= l_P/2$ or $M_P/2$ in full units. 

As a matter of fact, we believe that the relation (5.3) and the related idea of the {\it extremal entropy of 
elementary black hole} mentioned above underlie the {\it universality} of $S_S[site]$ in our present scheme, 
although our scheme has not determined, from its first principle, the {\it universal constant} itself, such like $S_S[site] 
= \pi\ (4.22)$ which was suggested from the Bekenstein-Hawking relation (4.21) or $\eta =1/4$ and on which our preliminary 
arguments in subsection 4.3 have been based.

2.  Let us turn our attention to ${\bar N}_S[site] (= 1/(4\mu_S M_S) see (4.28)),$ on which we have pointed out 
in subsection 4.3 also to have a certain universal nature as $S_S[site]$ on account of the relation (4.30). 
In accordance with the argument on $S_S[site]$ given above, one finds out that ${\bar N}_S[site]$ is nothing but 
the average number of $D_0$ brane inside the {\it elementary black hole} mentioned above, and its value is constrained 
to have a certain fixed value given in Eq. (4.37), that is, 
\begin{eqnarray}
{\bar N}_S[site](= 1/(4\mu_S M_S)) \sim 1/0.12.
\end{eqnarray} 
in accord with $S_S [site] = \pi $.

3. Next, let us consider Schwarzschild black holes with $R_S$ in general, other than the elementary black 
hole considered above. 

With respect to the entropy, Eq. (5.2) clearly tells  us that the entropy of the {\it elementary black hole}, 
$S_S[site]$ discussed above in 1., plays the role of the {\it element} of the entropy of black holes in general, 
equally constituting the entropy of the individual [site].    

With respect to the mass of black holes, let us remember Eq. (4.14) [AMR], which becomes  
\begin{eqnarray}
M_S (= M(V_3^{R_S})) = {\cal A}(V_3^{R_S})\ {\mu_S {\bar N}_S[site] \over 4 \pi} 
 = n_{dof}(V_3^{R_S})\ \mu_S {\bar N}_S[site],
\end{eqnarray}
when the system tends to schwarzschild black hole with radius $R_S.$ 

One notices that Eq. (5.5) combined with (4.28), nicely reproduces the relation (4.15), that is, 
\begin{eqnarray}
R_S = 2 M_S,
\end{eqnarray}
as expected, on account of $ {\cal A}(V_3^{R_S}) = 4\pi R_S^2. $

At this point, it is quite interesting to ask what happens in each [site] when the total system 
forms a black hole with radius $R_S ( >> l_P).$ As seen in (5.4), one should remember that 
${\bar N}_S[site]$ together with $S_S[site]$ must be {\it universal} and remains same as in the case of 
{\it elementary black hole}, while $\mu_S$ is subject to the relation Eq. (5.4), that is,
\begin{eqnarray}
\mu_S  (=  1/(4 {\bar N}_S[site] M_S) \sim 0.03/ M_S = 0.06/ R_S, 
\end{eqnarray}
in accord with $S_S[site]= \pi.$  Therefore, one finds that the effective mass at each site, i.e., 
$\mu_S {\bar N}_S[site]$ is clearly smaller than that of {\it elementary black hole} with $R_S =l_P$,
being insufficient for the formation of elementary black hole.
  
4. On the other hand, Eq.(5.7) tells us very importantly that the effective mass of $D_0$ brane inside black holes, 
$\mu_S$, is inversely proportional to the respective black hole mass $M_S$, almost in accord with the Hawking 
radiation temperature, $T_{H.R.} (= 1/(8\pi M_S) \sim 0.04/M_S),$ see (4.35)), showing the strong correlation between them,
 $ T_{H.R.}  \sim (0.04 /0.03) \mu_S$ (4.38). In fact, at the end of subsection 4.3, we have emphasized that 
Hawking radiation temperature $T_{H.R.}$, instead of $T_S$, plays the role of a genuine temperature of the black hole 
system under $[BHC]$, in connection with this strong correlation with $\mu_S$.

Motivataed  by this fact, we attempt in what follows, to examine further the physical implication of $\eta$, although 
we have so far supposed tacitly $\eta$ to be 1/4 according to Hawking. In fact, we try to introduce in our scheme an ansatz 
which enables us self-consistently to equate $\mu_S$ to $T_{H.R.}$ and leads us to $\eta$ slightly shifted from $1/4.$ 
As will be easily understood from the argument in 3., this possibility turns out to be possible only when the parameter 
$\eta$ in Eqs. (4.19) or (4.20), is treated as a free parameter, not necessarily fixed to be $1/4$ from the beginning, 
but rather to be decided self-consistently.

As a matter of fact, one finds out that the new ansatz is well brought in our present scheme, together with the relation 
(4.20):
\begin{eqnarray}
S_S[site] = 4 \pi \eta,
\end{eqnarray}
\begin{eqnarray}
[Ansatz]  \hspace{1cm} \mu_S = T_{H.R.} = {1 \over (8 \pi M_S)(4 \eta)}, 
\end{eqnarray}
taking into account $ T_{H.R}^{-1} = 8 M_S S_S[site]$ (4.33).

Noticing the relation 
\begin{eqnarray}
4\mu_S M_S (= {\bar N}_S [site]^{-1}) = (8\pi \eta)^{-1},
\end{eqnarray}
derived from (5.9), one finds out that the parameter $\eta$ is determined from the equation 
\begin{eqnarray}
\ln(1+ (8\pi \eta)) + 8\pi \eta \ln (1 + (8\pi \eta)^{-1}) ( = S_S[site], see\ (4.30))= 4\pi \eta,
\end{eqnarray}
with the result
\begin{eqnarray}
\eta \sim 0.22\ ( \sim 1/4  \times 0.88).
\end{eqnarray}
On the other hand, Eq. (5.9) gives rise to 
\begin{eqnarray}
\mu_S = T_{H.R.} = {1 \over (8 \pi M_S)} \times (0.88)^{-1}.
\end{eqnarray}

Needless to say, we are not granted to much emphasize the physical significance of the deviation factor $0.88$ in 
Eqs.(5.12) or (5.13) with respect to $\eta$ or $T_{H.R.}$, if we consider our simple and crude method to treat such 
a $D_0$ brane gas system under $[BHC]$. We simply expect that the above consideration serves for making clear 
the unknown physical implication of the parameter $\eta$ which possibly relates $\mu_S$ with $T_{H.R.}$ 
from the more profound level, calling to mind the Wien's displacement law in the black body radiation.
  
In conclusion, it should be emphasized again that almost all results derived in this paper, essentially reflect 
the kinematical holographic relation (3.1) or (3.12) in YSTA.  As was remarked in Introduction, the kinematical 
reduction of spatial degrees of freedom is expected to hold widely in the noncommutative space-time. So, it is 
interesting to examine how the kinematical reduction of spatial degrees of freedom may occur in the noncommutative 
space-time algebra other than the present YSTA extensively studied so far. This consideration will give us an important 
clue to seek for a candidate for the ultimate theory, which is expected to satisfy the kinematical holographic relation 
and to be free from UV- and IR-divergences.$^{[26],[33]}$

The above view  on the kinematical holographic relation in Yang's space-time reminds us of the following Bekenstein's view  $^{[9]} $ on the holographic principle, in addition to the remarks by Strominger $ ^{[12]} $ cited in Introduction: "This (holographic) principle is viewed as a guideline to the ultimate physical theory. A consistency requirement on it is that the boundary of any system should be able to encode as much information as required to enumerate and describe the quantum states of the bulk system. $ \cdots $ "   

Finally, with respect to the idea of {\it elementary black hole}, which has played a vital role in our scheme, 
giving the qualification of being {\it universal} to $S_S[site]$ or ${\bar N}_S[site]$ in connection with the Bekenstein 
parameter $\eta,$ it is interesting further to investigate its possible roles in the physical world.

\end{document}